*Research Paper*

# Voxel selection framework based on meta-heuristic search and mutual information in brain decoding

**Osama Hourani, Nasrollah Moghadam Charkari, Saeed Jalili**

Friday 31st August, 2018

**Abstract**

Visual stimulus decoding is an increasingly important challenge in neuroscience. The goal is to classify the activity patterns from the human brain; during the sighting of visual objects. The inputs are fMRI data, and the stimuli names are the outputs. One of the crucial problems in the brain decoder is the selecting informative voxels. Selection methods fall into one of three types: filter; wrapper; and embedded. We propose a meta-heuristic voxel selection framework to decode the brain activities of visual stimuli accurately. It is composed of four phases: pre-processing of fMRI data; filtering insignificant vox- els; post-processing; and meta-heuristic selection. The main contribution is benefiting a meta-heuristic search algorithm to guide a wrapper voxel selection. The main crite- rion to nominate a voxel is based on its mutual information with the provided stimulus label. The Support Vector Machine, with a linear kernel is operated as an efficient clas- sifier. The leave-one-subject-out cross-validation scenarios have been used to assess the generalization of proposed framework. The results show impressive accuracy rates which are 92.4% and 92% for DS105 and DS107 respectively. This outperforms the most of existing brain decoders in similar validation conditions. Brain decoding framework has been proposed, by focusing on the information quantity in fMRI data. SA, is a search al- gorithm that used in the wrappers methods to select informative voxels. The experimental results of proposed framework are very encouraging which can be successfully used in the brain-computer interface, BCI, systems.

*Keywords:* Brain Decoding, fMRI, Mutual Information, Simulated Annealing, Voxel Selection, Visual Stimulus.

# 1 Introduction

The study of brain decoding goes back to two decades and a half. Brain decoder is a computational model to establish a proper correlation between the brain activity and the outside stimulus. The first aim of a brain decoder is to classify neural activities into related high-level cognitive states such as the processing of a stimulus (viewed objects); emotions; and mental illnesses. The second one is localizing the brain regions that are strongly selective to a specific stimulus; this helps neurologists to understand the underlying brain activity. The attractive application of brain decoder is in designing the brain-computer interface, BCI, systems.

There are different neuroimaging modalities to record brain activities. For instance, Electroencephalography (EEG), Magnetoencephalography (MEG), Positron-emission tomography (PET), and functional magnetic resonance imaging (fMRI). fMRI records the brain activities via tracking oxygen consumption in the brain regions; which known as blood oxygen level dependent (BOLD) signals. Even though fMRI scanners produce noisy and high-dimensional images, it remains a favorable imaging modality for brain decoding issue, because it is noninvasive and provides detailed information on deep brain structure and activity [47]. In fMRI data, it is difficult to distinguish between neural patterns due to the spontaneous fluctuations, even if they are obtained from the same person [13].

The validation scheme is used to assess brain decoder for inter-subject case, leave-one-subject-out cross-validation schemes; The state-of-art frameworks to select informative voxels from the fMRI images have been presented in [5, 53, 57]. The experimental results indicate how the stability of brain decoders is increased while over-fitting avoided.

If a voxel leads to decode viewed stimuli accurately, it will be hypothesized as "informative". Identifying and selecting the informative voxels is used in decoding and localizing brain activities. While devising a brain decoder based on fMRI data of perceiving a visual stimulus, the following questions could be raised: to what extent we trust the voxels? How accurate is the brain decoder? Where are the locations of these selected voxels in the human cortex? How are they allocated on the cortex; distributed or centralized within the same subject or among subjects? The advantage of the proposed framework is its significant accuracy rate using multi-class SVM in brain decoding, which makes it applicable in the brain-computer interface, BCI. Furthermore, it helps neurologists to localize neural activity accurately.

In the literature, "subject" refers to a person who set in the fMRI scanner and do the brain scan; in order to determine which part of his/her brain is activated by external stimulus. In the study of vision in the human brain, the "object" denotes a visual stimulus such as a picture of one visual object; viewed to the subject multiple times according to the task design. "Run" is referred as the fMRI data acquired in one experimental session also "trail" has the same meaning, "session" may includes one or more trials. Task timing or task "onset" includes the time of object viewing to the participant during fMRI acquisition; starting and ending time. The "voxel" denotes one point in the three-dimensional space of brain image. Each voxel in fMRI images has three indices (X, Y, and Z), its values represent the BOLD signal across time.

Figure 1 shows the main phases of the proposed framework. The pre-processing and voxel filtering are performed by FSL software, but the post-processing and wrapper selection of voxels are implemented in Matlab environment. Hereafter, we call the proposed *Meta-heuristic Filter and Wrapper Voxel Selection* framework MFWVS in short. In the voxel filtering, the general linear model, GLM, filters the whole brain voxels by localizing activities, which are associated with the visual stimulus.The atlas information filter the statistical maps (GLM's outputs). Accordingly, we run an iterative procedure; to select the voxels with maximum likelihood relation to the external stimulus from each anatomical region. Then, we execute the voxels post-processing which consists of multiple processes; normalization, value scaling and discretization. Post-processing makes voxels appropriate to calculate mutual information and learn the classification model.

The main contribution of this paper is using the Simulated Annealing algorithm, SA, in wrapper voxel selection approach, to find best voxel subset in the input voxel space. SA algorithm searched for the optimal values of ($α, β$) parameters in our voxel scoring algorithm, mutual information voxel scoring, MIVS. The wrapper voxel selection helps the learning model to give a high accuracy rate. Further more, We apply voxel filtering to reduce the number of input voxels to tackle the time-complexity problem in MIVS. The proper selection of the mentioned parameters leads to the finest possible accuracy rate. Finally, visualization of the selected voxels in within-subject and between-subjects scenarios gives evident insight into underlying brain activity. This paper is divided into five sections. In section two, we discuss the state-of-the-art studies on brain decoding and voxel selection. Section three presents the proposed framework in detail. Section four, shows the significant results of the proposed framework and discuss it with the recent studies in the field. The final section presents the concluding remarks and declares the important points of the proposed framework.

## 2 Literature Review

Brain decoding has become a hot area of neuroscience research in recent years. The role of brain decoder is to discriminate between external stimuli depending on some special patterns in brain activities. Furthermore, several surveys have discoursed the brain decoding problem especially from visual object perspective [1, 4, 18, 20, 38]. There are three approaches to solve brain decoding problem: neural pattern classification, stim- uli identification, and reconstruction [36]. In this study, we only discuss the neural pat- tern classification which can be assorted into four different approaches; nonlinear such as multi-layer perceptron [12, 17, 50], probabilistic methods [5, 15, 35–37, 46], graph- based decoder [39, 43, 44] and multivariate pattern analysis (MVPA). MVPA may in- cludes some learners like Support Vector Machine (SVM) linear discriminant classifiers, and k-nearest-neighbor [8, 10, 32]. In general, brain-decoder is composed of three major steps: pre-processing, voxel (feature) selection, and model (decoder) learning. Because of the large volume of data, using some specific software tools would be crucial. In this regard, some of the most popular pre-processing software like FSL [23, 49, 54] and SPM [14, 41] have been developed.

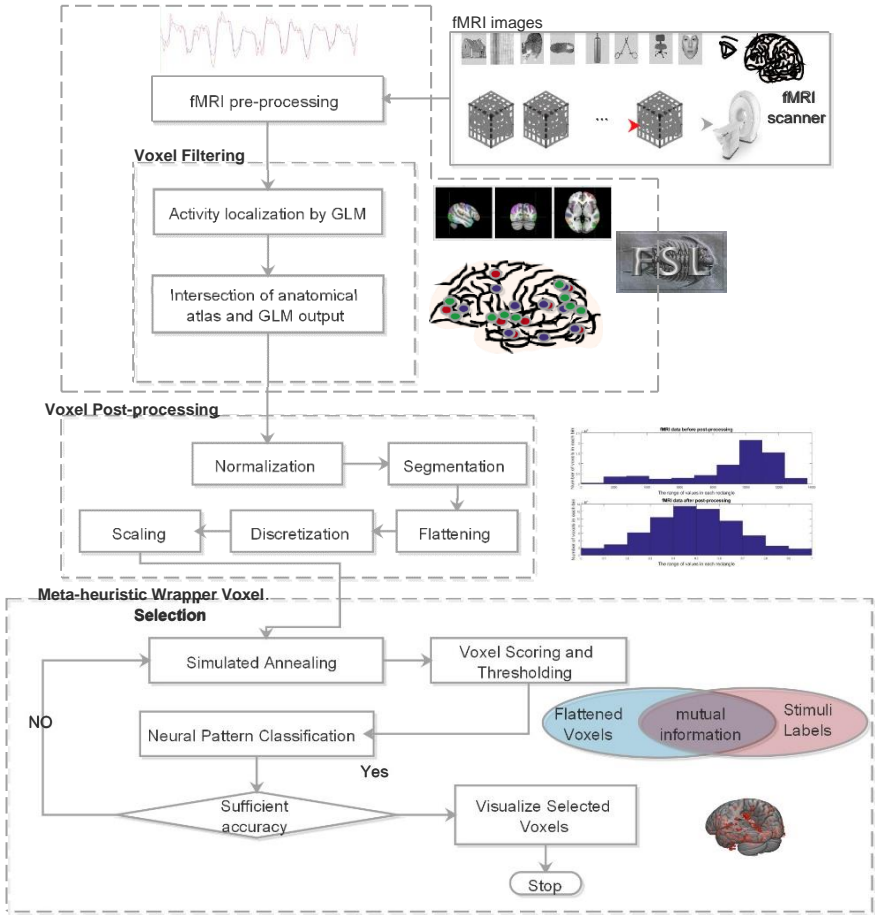

FIGURE 1: The proposed framework for meta-heuristic voxel selection based on mutual information

Principally, feature selection is categorized into three approaches; filtering, wrapper and embedded feature selection [45]. Filtering approach calculates some measures between features and labels such as correlation, mutual information, and causality. The measures acting as scores to nominate the important features subset. Filtering methods work independently of the learning process. Wrapper feature selection approach generates a subset of features and accesses its accuracy rate from learning model. Wrapper method depends on the learning process to evaluate the selected voxels (features) subset. The wrappers are divided into deterministic such as sequential forward or backward selection, SFS or SBS, and meta-heuristic search algorithms like simulated annealing, hill climbing [27, 45]. The embedded feature selection works similar to wrapper methods, but it performs the feature selection with learning process jointly such as decision tree and elastic nets; some embedded feature selection methods introduced in [25, 32].

fMRI images are massive and high dimensional, so the voxel (feature) selection method plays an essential role in decoder performance. So, several early studies focused on efficient voxel selection methods. GLM is one of the famous voxel selection methods or neural activity localizers which are used for voxel selection in fMRI. It finds the most suitable voxels to the task model design; when the specific stimulus is provided [8, 19]. The studies [1, 5, 30, 33, 52] employed mutual information between voxels and the stimulus as voxel selection on activity localizing. In other studies Principal Component Analysis (PCA) has been used for voxel scoring. PCA method is based on variance analysis of time-series voxels; it finds the orthogonal components with high discrimination ability using a non-linear transformation [17, 31, 40]. In other studies, Graph-based methods have been used to localize brain activity and select the most important voxels [26, 39, 43]. As we try to propose a simple and fast voxel selection framework we use two levels of selection inspired by the previous studies. First the filter approach which is settled on statistical model and atlas information, this reduces the voxels number (dimensional space) and time-complexity for the next level. Second is wrapper, consists of voxel scoring algorithm to choose the informative voxel (feature) subset, and the Simulated Annealing search algorithm.

## 3 Materials and Methods

This section introduces the proposed framework for voxel-selection to localize the brain activities of viewing visual stimuli and decode these stimuli precisely. To find the most informative voxels, four major phases that are pre-processing, filtering, post-processing, and wrapper selection are conducted. In the following, each phase is discussed in detail.

### 3.1 fMRI datasets

In this paper, two real fMRI datasets are processed, including object viewing task in the human brain. The first one is a widely used fMRI benchmark for studying visual object recognition task in the human brain [19]. The second one consists of fMRI data related to viewing four categories [9]. We download these data from "openfmri.org" website, which has been deprecated to "openneuro.org". Table 1 shows the summary statistics of these datasets.

TABLE 1
Summary statistics of the used fMRI datasets

| Dataset Name | Subjects Number | Runs Number | Time points Number | Corrupted Subject | Objects Stimuli |
|---|---|---|---|---|---|
| DS105 | 6 | 12 | 9 | subject 5 run 8 | Faces, Houses, Cats, Bottles, Scissors, Shoes, Chairs and Scrambled |
| DS107 | 49 | 2 | 7 | subject 41 run 2 | Words, Objects, Scrambled Objects, Consonant strings |

### 3.2 fMRI pre-processing

Pre-processing of fMRI images was accomplished using FEAT (FMRI Expert Analysis Tool) Version 6.00, part of FSL library [23]. The Details of pre-processing steps are present in Appendix A.

## 3.3 Voxel filtering

The goal of this phase is to decrease the voxels number (feature space size) by concentrating on the most important cortical regions instead of whole brain structure. In the initial case, the number of voxels is so large; especially when the voxel size is converted to 2mm standard space. As a result, using an anatomical mask or atlas is highly recommended. In this regard, we suggest multiple steps to filter unimportant voxels. We only focus on the cortical regions of the brain and neglect others like cerebellum or sub-cortical structures; so we choose Harvard Oxford Cortical atlas. If we select all voxels in the atlas, the learning algorithm will face with a large feature vector. Thus, we use the intersection between atlas and the output of regression model. More information on voxel filtering steps are declared in Appendix B.

## 3.4 Voxel post-processing

Voxel post-processing is an essential phase in the proposed framework. Accordingly, the voxels data are standardized and arranged in a vector form to further use in learning process. The main steps are normalization, segmentation, flattening, discretization, and scaling. Let's denote the whole time-series of data could be expressed as a matrix where the columns represent as voxels, and rows indicate the time points of the related voxels. The above mentioned steps are applied on this matrix as follows:

- Normalization which adjusts values in a specific range for each column separately.

- Calculate the average of the time-series voxel in each column.

- Subtract the above average from each row "time point" for each column; this task leads to standardization and uniformity of voxels.

Segmentation of the time-series is the second part in post-processing. It determines what rows are related to specific stimulus in the time-series. It is committed to the timing values in the onset files which comes with datasets. For example, in DS105 each stimulus is presented during 9 time points (each time point is 2.5 seconds, TR=2.5s). Similarly, each stimulus is presented in 7 time points (each time point 3 second, TR=3s) for D107. Flattening is responsible to reshape the time-series segment into one vector. The discretization plays an important role in reducing the noise, also eliminates the fluctuations in values. Finally, scaling projects the values of voxels between 0 and 1, this procedure helps the SVM model to learn faster. Figure 2 compares the histogram of voxels magnitudes before and after post-processing. X-axis presents the bins of a particular magnitude range, Y-axis indicates the number of voxels in the range of related box. Before post-processing, the magnitudes are varied from 0 to 14000 which has a negative effect on the classification performance. Figure 2-a, shows total characteristics of data before post-processing. As it is shown the majority of voxels fall into the range of 8000 to 12000. Refer to Figure 2-b, the voxels have become similar to a normal distribution after post-processing.

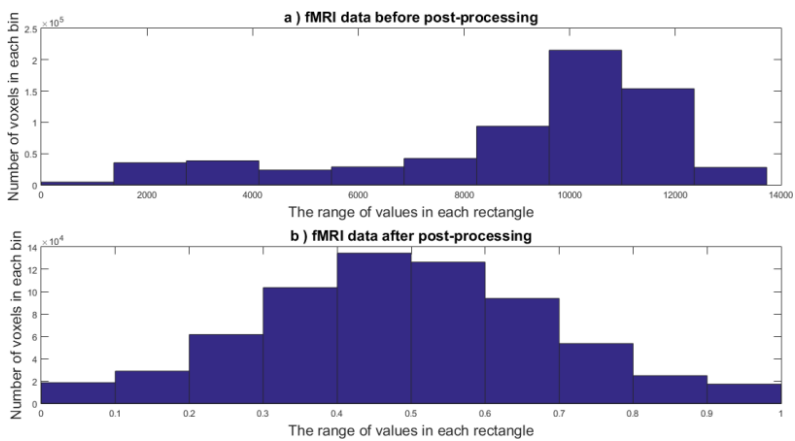

FIGURE 2: Histogram of voxels magnitudes before and after post-processing

## 3.5 Meta-heuristic wrapper voxel selection

Voxel selection is essential in building the brain decoder, due to high number of voxels in the brain images. Similar to feature selection the voxels selection methods fall into one of three types filter, wrapper (meta-heuristic or deterministic) and embedded. Wrapper approach embed the model learning within the feature subset search [34, 45]. We choose wrapper method because it is suitable to tune the both classification and scoring algorithms. At first, we implement a special algorithm to calculate the voxel scores based on mutual information, and select the important subset. Then, simulated annealing is employed to find the best voxel subsets according to the parameters in a mutual information algorithm. Therefore, the proposed framework recognized as meta-heuristic. Figure 3 illustrates the arrangement of processes in voxel selection framework, in this case each fold is related to one subject.

### 3.5.1 Simulated annealing

Simulated Annealing, SA, is a probabilistic and memory-less search method. When solving a specified function, SA is used to find a proper approximation of the global solution. Multiple versions of approximation algorithms are used in the wrapper voxel selection approach [2, 6, 11, 28, 51]. Definitely, SA is often used when the search space is dis- crete such as traveler salesperson problem. Adaptive Simulated Annealing, ASA, is also a global optimization method, which inspects the search space more efficiently than other previous simulated annealing algorithms [21]. In the proposed framework, ASA is used to investigate the global optimal values of two parameters $a$ and $\beta$ in the mutual information algorithm which aim to minimize the error rate of the SVM classification algorithm. The error rate is defined as below:

$$ErrorRate = \frac{\text{number of false classified instances}}{\text{number of all classified instances}} \quad (1)$$

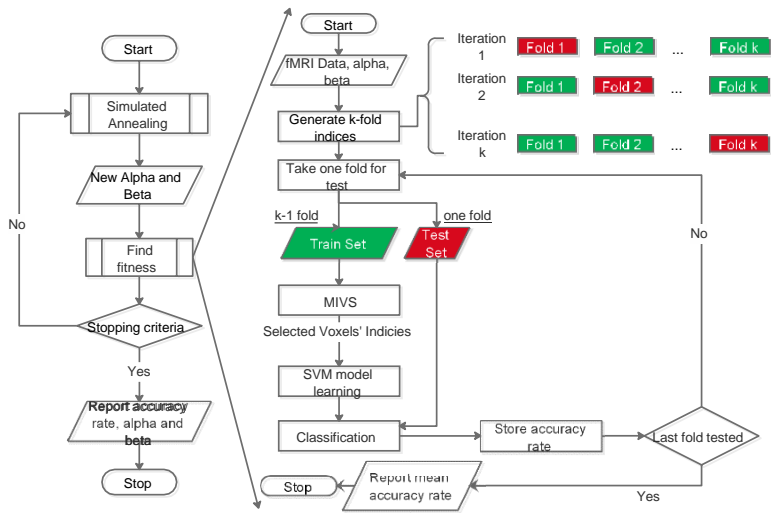

FIGURE 3: Wrapper voxel selection approach in inter-subject cross-validation scheme

The error rate of the SVM classification model is acting as fitness function, and the ASA tries to minimize it, by searching the space of $\alpha$ and $\beta$ values. The default parameters are used in ASA algorithm such as annealing fast, exponential temperature, and SA acceptance function. We determined the start solution of alpha and beta as $(0.003, 0.56)$. It is necessary to mention that lower and upper bounds of two parameters are very important to guide the whole search.

### 3.5.2 Voxel scoring

Mutual information, MI, is a measure to estimate the statistical dependency between two numeric phenomena [7, 16]. Mutual information between two discrete random variables is calculated as shown as follow:

$$MI(A, B) = \sum_{a \in A} \sum_{b \in B} p(a,b) \cdot \log\left(\frac{p(a,b)}{p(a) \cdot p(b)}\right) \qquad (2)$$

The mutual information between two independent variables is presented as $MI(A, B)$. The term $p(a, b)$ is the joint probability distribution function for A and B. $p(a)$ and $p(b)$ are the marginal probability distribution functions of A and B, respectively [7]. The mutual information voxel scoring, MIVS, finds the importance score of the voxels based on its mutual information with the viewed-object labels. Figure 4 shows the flowchart of the voxel selection algorithm by mutual information scoring. The inputs of MIVS algorithm are:

- high-level labels or visual objects such as house, scramble, cat, shoe; and so forth. For each label, an integer value is assigned. For instance, house→1, scramble → 2.

- Flattened BOLD signal values of voxels in the tabulated form $X$, each row presents one block of the signal, which is relevant to one high-level label.

- The parameter $a$ is the threshold value for the accepted mutual information. If a voxel which has a mutual information value above the threshold, it will be selected.

- The parameter $\beta$ is used as a threshold value on a BOLD signal. In other words, this parameter determines which voxel is activated and when?

The major instructions of the MIVS algorithm are described in Figure 4. First, the labels for each instance were converted to the binary form as follow $1 = (10000000)$, $2 = (010000000)\ldots 8 = (00000001)$. Then, the threshold $\beta$ would be employed on the voxel values. This threshold determines whether a voxel is triggered or not. MI vector for each label is calculated. It consists MI between every voxel and the binary label. Then, the summation of the MI vectors for all classes is found. The algorithm returns the indices of the most important voxels after applying $a$ threshold on final mutual information values of each voxel. The loop bounds of two nested for-loops in the flowchart are defined as label-number, k, and voxel-number, n, respectively. If we assume the time complexity of mutual information calculation becomes $O(m)$ where m indicates the number of time points in each voxel, then the overall time complexity of the algorithm will be $O(k \times n \times m) \; \dot{} \; O(n^3)$ where $k, m$ $n$; the number of overall features is calculated as following $n = $ (voxels number × time points number). We hold this issue (high dimensions) by adding the voxel filtering phase; which discussed in the section 3.3.

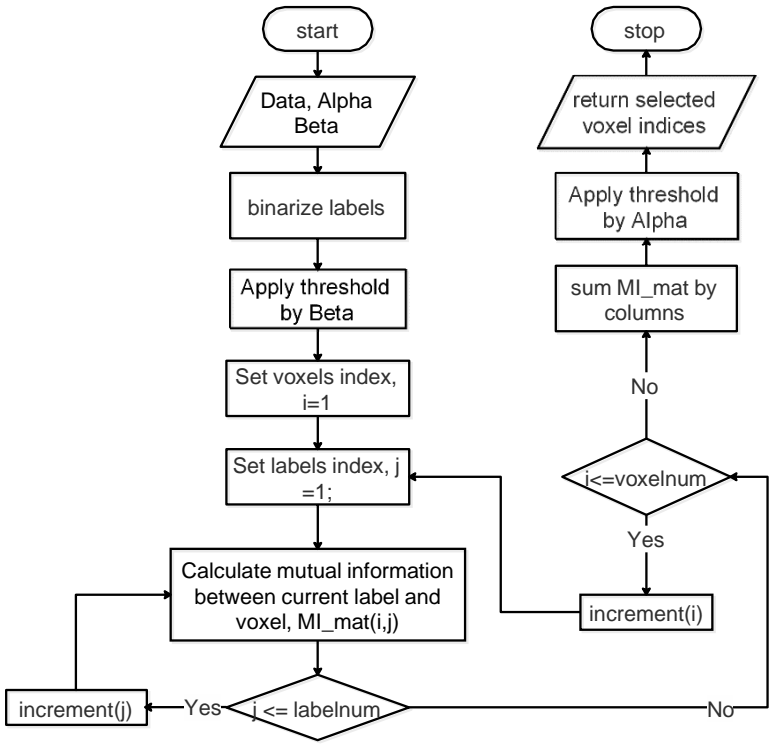

FIGURE 4: Flowchart of voxel scoring algorithm by mutual information

### 3.5.3 Neural pattern classification

The objective of a brain decoder is to distinguish between neural patterns and recog- nize the related high-level cognitive state, similar to the classification algorithms. In this work, we have groups of time-series voxels of the stimulated brain; each group can be categorized into one of the multiple visual objects. For example, in the dataset DS105, there are eight different classes (visual objects), and four objects in dataset DS107. In this regard, the brain decoder acknowledged as a supervised classification problem. The wrapper methods works iteratively until find the best voxel subset. The accuracy of the learning process in each iteration fed into the SA algorithm as fitness value. In the learning process, different models can be used such as SVM, logistic regression and decision tree [5, 8, 43]. Support Vector Machine, SVM, is a supervised learning model used for both classification and regression problems. The original version of SVM was initially introduced for binary classification.

We have employed the LIBSVM [3], since the multi-class is the main concern in this work, LIBSVM converts the SVM binary models to the multi-class classifier, by utilizing a one-versus-one scheme; In our investigation, it has been shown that the linear kernel surpasses other ones in SVM for intra-subject scenario; However, in the inter-subject, the RBF kernel generates accurate results in comparison with linear and polynomial kernels.

## 4  Results and Discussion

Two real fMRI datasets are investigated [9, 19]. Visual object viewing in the human brain is the examined task. FSL library tools carried out the pre-processing and first level GLM analysis of fMRI images. The learning and voxel scoring algorithms have been executed in Matlab environment.

### 4.1  Evaluation measure and scheme

Accuracy is a description of systematic errors, a quantity of statistical bias, as this reason a change between the result labels and the reference values, also called as trueness [42]. The overall accuracy rate is calculated as below:

$$AccuracyRate = \frac{\text{number of correct classified instances}}{\text{number of all classified instances}} \quad (3)$$

The validation approach is leave-one-subject-out, it is used to evaluate the inter-subject variabilities. This approach is beneficial in evaluating the mind-reading models because it is similar to nature. Naturally, the model trained data comes from multiple persons. The model will be tested with unseen neural data, which came from a new subject (person). This style is valuable for generalization of the brain decoder across multiple subjects. In this scenario, all the samples of one participant put aside for testing and the date from other participants, are used for learning. Then, the cross-validation procedure is repeated for the entire participant, each fold is related to one subject Figure 3.

## 4.2 Results and Anatomical analysis

In DS105, the mean accuracy rate in inter-subjects validation scenario is 92.4%. The median accuracy rates are the same for all kernel types. More than half of accuracy rates are above 91%. The RBF and linear kernels give similar performances. While in DS107 the mean accuracy rate is 92% and the median is about 91.67% in linear kernel, it shows more stability than RBF and Polynomial kernels because its accuracy has lower standard deviation.

TABLE 2

anatomical locations of the largest blobs of voxels in two datasets

| Dataset Name | Brain Region | Percent | Stimulus Name |
|---|---|---|---|
| DS105 | Left Lateral Occipital Cortex, inferior division | 6.07 | Scramble |
|  |  | 7.34 | Shoe |
|  | Left Paracingulate Gyrus | 4.34 | House |
|  | Left Superior Parietal Lobule | 4.62 | Bottle |
|  |  | 6.66 | Cat |
|  | Right Lateral Occipital Cortex, inferior division | 5.31 | Scissors |
|  |  | 6.30 | Chair |
|  | Right Temporal Occipital Fusiform Cortex | 5.65 | Face |
| DS105 | Right Occipital Pole | 4.75 | Consonant String |
|  |  | 4.21 | Objects |
|  |  | 4.64 | Words |
|  | Right Superior Temporal Gyrus, posterior division | 3.73 | Scrambled Objects |

Figure 5 presents a clear distribution of voxels in LOC and TOF for DS105. As it can be found, the spread of voxels in the Occipital Pole and Superior Temporal Gyrus are obvious to DS107. Table 2 gives a detailed explanation of the anatomical regions in Figure 5.

## 4.3 Comparison with state-of-the-art methods

The proposed framework provides the "informative voxels" to the classification model which leads to high accuracy rates Table 3 compares the performance of different voxel selection and classification methods for all subjects of both datasets. The comparison criteria are the number of selected voxels (without considering the number of time points in each voxel), selection method, and accuracy rate. All rows in the Table 3 are inspired from [57] but the main concern of comparison is the voxel selection method.

Hence, studies in Table 3 have employed voxels with more than 3 mm in size, while our framework utilizes voxels with 2mm size in the MNI152 standard space. On the other hand, the other methods are established on Haxby ROIs; the Ventral Stream in the human brain. Instead, MFWVS employs anatomical information derived from Harvard Oxford Atlas, which includes 96 cortical regions. It is worthy to mention that the major difference between the proposed method and the state of methods is in the voxel selection strategy, which is crucial in recognizing the visual activity of the human brain. The accuracy rates differences between proposed framework and other methods are highlighted Table 3. The results, as shown in Table 3, indicate that proposed framework states a superior accuracy rate in most cases, except one.

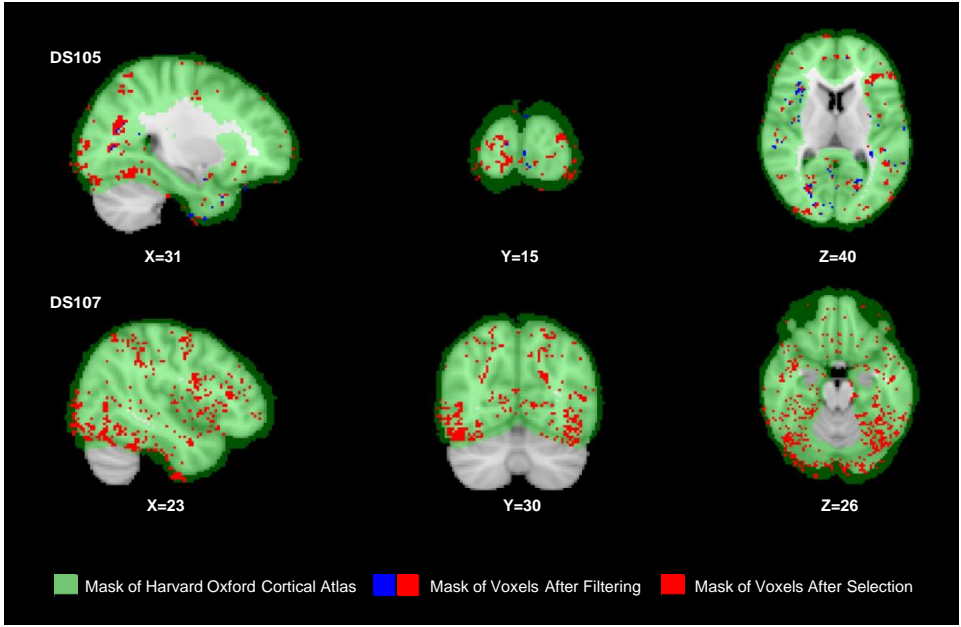

FIGURE 5: Selected voxels locations in 3D image per stimulus

To investigate the significance of accuracy rates in the proposed framework, a two-way ANOVA test has been performed. We assume that the accuracy variables have a normal distribution. For each method, the mean and std are used to generate 10000 variables. Then, the generated variables from all subjects are put together to form one variable for each framework results. The test indicates that MFWVS report has superior accuracy rates for DS105 at significance level p=0.001. Furthermore, MFWVS exceeds all other methods except [57] for DS107 with one percent variation as shown in Table 3.

The number of selected voxels have not been reported in the mentioned references, so we did not put them in the Table 3 for comparison. But in the previous work of Yousefnezhad [56] the number of selected voxels are varied between 100 and 1200.

TABLE 3
Comparison of accuracy rates for inter-subjects classification results in both datasets

| Method | Voxel selection type | Voxel selection details | Classification method | Dataset Name | |
|---|---|---|---|---|---|
| | | | | DS105 | DS107 |
| [8] | Filter | First extraction of brain tissue. One way ANOVA test, correlation analysis | multi-class SVM, SVD, Linear discriminant | 18.3 ± 4.07 | 38.01±2.56 |
| [29] | Filter | The intersection between Desai Maximum Probability atlas and GLM functional activity localizer | SVM, soft-margin classifier | 28.72±2.37 | 68.51±1.07 |
| [39] | Embedded | Specific functional ROIs, voxels with strong anatomical links selected | Graph based classifier | 50.61±4.83 | 89.69±2.32 |
| [57] | Filter | Anatomical Pattern analysis | Adaboost multi-class | 59.21±2.05 | 95.61±1.83 |
| MFWVS (Proposed) | Filter, Wrapper | Both filter and wrapper methods with simulated annealing | multi-class SVM | 92.4 ± 9.18 | 92 ± 8.3 |

## 5 Concluding Remarks

This paper provided a novel framework, MFWVS, to discover the informative voxels, related to any fMRI task-based dataset. Even neurologists or engineers can utilize the MFWVS for neural patterns decoding. Each process in the framework acts as a piece of mosaic panel. Missing one piece leads to undesired decoding accuracy rate. This framework has been tested on two popular fMRI datasets, of object viewing tasks, DS105 and DS107. MIVS algorithm takes more time to run, especially in the inter-subjects case. In this regard, reducing the voxel space is highly recommended before starting wrapper voxel selection. The reduction of the voxel space is accomplished in voxel filtering phase. This task was done by using the localizer's outputs with anatomical atlas. Multi-level voxel filtering and selection reduced time-complexity of the whole voxel selection process. Since the proposed framework depended on the GLM output, this caused some restrictions as discussed in [55]. Moreover, if the task timing information about the test fMRI images have not supplied, applying MFWVS will be impractical. Voxel post-processing performed extra refinement on the voxels values and led to improving the decoder performance. At the present, no specific rule for determining the main parameters in MIVS, $\alpha$ and $\beta$, has been found. Indeed, these parameters were highly subject correlated. We have employed Simulated Annealing optimization to find the best values of $(\alpha, \beta)$ to solve this issue we estimated these parameters; the average of them acquired from this study. The visualization of selected voxels into 3D brain images helps to understand the underlying brain activity and this gives a high advantage to the neurologists. The anatomical locations of dense voxels are consistent with anatomical findings in the literature, LOC, TOF as shown in related tables such as Table 2. The density or sparsity of voxels can be noticed clearly by expertise or visualization software like "atlasquery" (provided with FSL library) can provide detailed anatomical analysis. The reported accuracy rates by MFWVS framework are very promising, which supported our hypothesis on informative voxels. The experimental results on two popular fMRI datasets indicated 92.4% in DS105 92% for DS107. The results were very interesting as the data are noisy and high-dimensional. Finally, how the brain patterns are varied? To answer this question the future work attempts to analyze the connectivity patterns between the selected voxel.

Meta Heuristic Voxel Selection 17

## Appendix A: fMRI pre-processing

Before starting FEAT first-level analysis, brain extraction and registration (into 2mm MNI152 standard space) are applied on all the fMRI data files. Afterward, the following processes (Feat 1st level analysis) are done Figure 6; non-brain structure removal, such as skull, neck and eyes, using BET [48]; motion correction using MCFLIRT; spatial smoothing using a Gaussian kernel of FWHM 5mm; grand-mean intensity normalization of the entire 4D dataset by a single multiplicative factor; high-pass temporal filtering (Gaussian-weighted least-squares straight line fitting, with sigma=18.0s). The FLIRT achieves the registration to structural MNI152 standard space with 2mm resolution. The registration algorithm applies normal search in 12 degrees of freedom (DOF) [22, 24].

To automate and increase the speed of fMRI processing using FSL, a collection of UNIX shell scripts have been implemented. These can operated as general, elastic and configurable toolbox for all datasets in openneuro website.

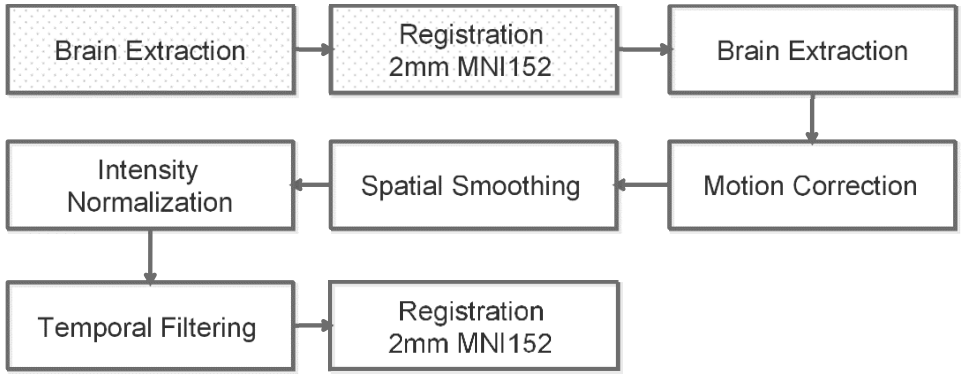

FIGURE 6: The pre-processing steps in the proposed framework

## Appendix B: Voxel filtering details

In summary, cortical regions and GLM statistical maps are our prior knowledge to achieve this task. The intersection between them guarantees to select the most important voxels in each anatomical region.

**Activity localization**

GLM is a statistical brain activity localizer, which is a soft generalization of linear regression to predict the relationship between two scalar variables; the brain activity and outside stimulus. The basic GLM is denoted by the following equation, $Y = XB + e$, where Y is the observed time-series fMRI data. X is the matrix of "regressors" which is indicated as the design matrix and B contains the parameters to be expected. In our study, we assume one regressor for each stimulus type in contrast with other stimuli. Error e is assumed to have a normal distribution. Time-series statistical analysis was achieved utilizing FILM with local autocorrelation correction [55]. GLM model produces statistic z-score maps which are 3D brain images, these images are used to calculate intersections in the next step.

**The intersection of atlas and localizer**

The activity related to each object is not only distributed in a predefined area but also in one-third of the whole brain. On the other hand, it is not beneficial to use the whole mask of the statistical z-score map due to the processing complexity. Specifying the most effective parts is not easy and straightforward. To tackle this problem, we attempt to extract the voxels with the maximum z-score in each region of Harvard Oxford Cortical atlas. This may ensure that much of the information would be kept by selecting a voxel for each stimulus from every ROI in the atlas. The criterion to filter a voxel is not only its anatomical location but also its similarity to the specific activity patterns. No threshold has been applied on z-scores in each anatomical region. In practice, all values which are below the maximum z-score will be neglected. This means that the remained voxels might be changed in each run and even in the same stimulus or in different subjects. The most frequent voxels tend to be more informative than the others.